%% file: main_latest.tex
\documentclass[letterpaper, 10 pt, conference]{ieeeconf}  

\IEEEoverridecommandlockouts                              
\title{\LARGE \bf
Truthful Production Uncertainty in Electricity Markets: A Two-Stage Mechanism
}

\author{Shobhit Singhal, Lesia Mitridati and Licio Romao
\thanks{The authors are with the Department of Wind and Energy Systems, Technical University of Denmark, 2800 Kgs. Lyngby, Denmark. Email: \{\tt\small shosi, licio, lemitri\}@dtu.dk}}

\usepackage{amsmath}
\usepackage{amsfonts}
\usepackage{subcaption}
\usepackage{graphicx}
\usepackage{acro}
\newcommand{\added}[2][]{#2}
\newcommand{\deleted}[2][]{}

\usepackage{tikz,pgfplots}
\pgfplotsset{compat=1.18}
\usetikzlibrary{shapes.geometric, shapes.arrows, arrows.meta, calc, positioning, fit, pgfplots.groupplots, backgrounds}
\usepgfplotslibrary{fillbetween,statistics}

\newtheorem{theorem}{Theorem}
\newtheorem{definition}{Definition}

\include{acronyms}

\begin{document}

\maketitle
\thispagestyle{empty}
\pagestyle{empty}

\begin{abstract}
    Renewable power sources have low production costs, but may result in high balancing costs due to the inherent production uncertainty. Current day-ahead markets elicit only point production profiles and neglect the degree of uncertainty associated with each generating asset, preventing the market operator from accounting for balancing costs in day-ahead dispatch and ancillary service procurement. This increases the total system cost and undermines market efficiency, especially in renewable-heavy power systems. To address this, we propose a new market clearing paradigm based on a two-stage mechanism, where producers report their production forecast distribution in the day-ahead stage, followed by the realized production in the real-time stage. By extending the Vickrey-Clarke-Groves (VCG) payments to the two-stage setting, we show appealing properties in terms of incentive compatibility and individual rationality. An electricity market case study validates the theoretical claims, illustrating the effectiveness of the proposed mechanism in reducing system cost.
\end{abstract}

\section{Introduction}\label{sec:intro}
Due to the negative climate impact of carbon emissions from fossil fuels, the world is moving towards cleaner sources of power, such as wind and solar energy. Such weather-dependent sources are hard to predict and cannot be fully controlled, and are therefore referred to as \emph{non-dispatchable}. Electricity markets dispatch resources ahead of delivery time, as conventional assets require advance operational planning and system operators procure ancillary services in advance to manage potential deviations from scheduled generation. Thus, in spite of a low cost of generation relative to dispatchable assets, a non-dispatchable asset may result in high ancillary service costs due to inherent uncertainty. This highlights a tradeoff between generation cost and uncertainty, to be considered in dispatch decisions. In power systems with a high share of non-dispatchable resources, cost effective dispatch must minimize total system cost, including both generation and ancillary service costs~\cite{bjorndal2018challenge}. Current electricity markets do not consider this tradeoff and optimize only for generation costs.

\definecolor{purple}{RGB}{170,51,119}
\definecolor{red}{RGB}{238,102,119}
\definecolor{green}{RGB}{34,136,151}
\definecolor{cyan}{RGB}{102,204,138}
\definecolor{yellow}{RGB}{204,187,68}

\tikzset{
node1/.style = {
        draw=red!60,
        line width=1pt,
        rounded corners,
        fill=red!20
    },
arrow/.style={
-{Latex[length=2mm]},
line width=0.4mm,
}
}

To address this, previous works consider a two-stage stochastic dispatch problem, where the market operator optimizes for the expected system cost including generation costs of producers, and ancillary service procurement and expected activation costs~\cite{morales2012pools,pritchard2010single}. The day-ahead and real-time prices are given by the dual variable of the respective power balance constraints. The resulting mechanism is revenue adequate and recovers producers' cost, in expectation, and produces a consistent day-ahead price equal to the expected real-time price~\cite{zavala2017consistent}, while \cite{kazempour2018scenario} extends the mechanism to satisfy revenue adequacy and cost recovery by scenario, albeit sacrificing cost efficiency. Since implementing the two-stage stochastic dispatch problem can be computationally prohibitive, several authors have proposed a single-stage approximation of the two-stage stochastic ideal, which requires fewer structural changes in the current market process. For instance, authors in~\cite{mays2021quasi} advocate for including an operating reserve demand curve (ORDC) in the day-ahead dispatch problem, representing the expected reserve activation cost,~\cite{morales2014improved} computes an optimal cap on the dispatch volume of the non-dispatchable producers, and~\cite{dvorkin2020chance} models the required amount of reserve procurement through a chance-constraint. 


A major assumption in the above works is the public availability of an asset's production uncertainty. While weather forecasts are widely available, the production behavior of individual assets, especially hybrid plants that combine renewables with storage or power-to-X technologies, depends on specific plant characteristics and operational policies known to the asset owner, preventing a system operator from fully inferring the underlying production uncertainty~\cite{dvorkin2019asymmetry}. Current electricity markets do not elicit uncertainty information but only point production profiles. In this paper, we consider a setting where production uncertainty is reported by the market participants, shifting the responsibility of characterizing an asset's uncertainty from the system operator to the asset owners, who have better access to this information.


However, truthful reporting is not always in a producer's interest: an efficient dispatch is more likely to leave a highly uncertain producer undispatched, creating an incentive to underreport uncertainty. A few works have considered reporting of probabilistic production forecast in the day-ahead market. For instance, \cite{tang2015random} develops a \ac{VCG} inspired two-stage mechanism that incentivizes \acp{WPP} to report true forecast distribution, assuming that they can not alter their realized production in the real-time stage. This is not realistic, as renewable producers can often curtail their output.
\added{\cite{papakonstantinou2016information} prices imbalance settlement using scoring rules that measure the deviation of production from the reported forecast distribution, recovering the two-price scheme for point forecasts. However, it is not clear whether this pricing rule is incentive compatible. In contrast, \cite{koumpis2026restoring} does not elicit forecast distribution, but evaluates the scoring rule against a reference distribution estimated from historical data. This requires a stable participant population over time, whereas our framework is independent of historical data. Authors in~\cite{hubner2026stochastic} take a different approach, where they trade state-contingent contracts at the day-ahead stage, clearing market for all possible uncertainty realizations ahead of time, leaving no imbalance to be settled in real-time. However, the number of states needed to describe the uncertainty grows exponentially with the number of stochastic participants, and incentive compatibility relies on a large market with price-taking participants.}

\added{In this paper, we address the above gaps by designing a two-stage stochastic electricity market (illustrated in Fig.~\ref{fig:twostagestochmarket}) based on the setting introduced in~\cite{ieong2007stochastic}, where the market participants report a production forecast distribution in the first stage and the realized production in the second stage. The market operator optimizes for the system cost at each stage, where the stage payments are designed to incentivize reporting of truthful information at each stage. Specifically, our contributions are as follows:}
\begin{enumerate}
    \item \added{We instantiate the framework of~\cite{ieong2007stochastic} as a two-stage stochastic electricity market that accounts for the total system cost in dispatch and reserve decisions, with stage payments incentivizing truthful reporting of asset characteristics including production forecast distribution.}
    \item \added{We extend the payments in~\cite{ieong2007stochastic} to be individually rational, and analytically show that the mechanism is sequentially ex-post incentive-compatible and ex-ante individually rational.}
    \item \added{We numerically evaluate the proposed mechanism using an electricity market case study, validating our theoretical results. A key result is that producers' payments decrease with increasing production uncertainty, while no producer can gain by unilaterally misreporting their degree of uncertainty.}
\end{enumerate}

The paper is organized as follows. Section~\ref{sec:twostagemarketsetting} introduces the two-stage stochastic market setting, while Section~\ref{sec:stochasticdispatch} describes the market operator's decision problems and Section~\ref{sec:mechd} defines the payment rules. Section~\ref{sec:numexp} describes the numerical experiments, and Section~\ref{sec:conc} concludes the paper.

\section{Two-Stage Stochastic Market Setting}\label{sec:twostagemarketsetting}

\added{Consider a collection of $n$ assets participating in an electricity market with a time-ahead stage (analogous to the day-ahead market) and a real-time stage (analogous to imbalance settlement). Let $\mathcal{N}:= \{1,\dots, n\}$ denote the set of participants. Then, for each participant $i\in\mathcal{N}$, let $\theta_i\in\Theta\subset\realN{p}$ denote $p$ asset characteristics, such as power generation and cost parameters, referred to as their \emph{type}. Current market structures require participants to report a single deterministic type at the time-ahead stage, e.g., their production capacity. This poses a challenge for many assets that observe their type at the real-time stage and have access only to a forecast distribution at the time-ahead stage; for instance, \acp{WPP} know only probabilistic forecast of their production capacity. Formally, let $q_i\in\mathbb{Q}$ denote a forecast distribution of participant $i$'s type, where $\mathbb{Q}:=\Delta(\Theta)$ is the set of probability distributions over $\Theta$. In contrast, we envision that the market operator elicits type distributions $q_{\mathcal{N}}$ at the first stage, and determines first-stage decisions $x\in\mathcal{X}_\mathcal{N}$ such as dispatch and reserve procurement, where $q_{\mathcal{N}}:= (q_i)_{i\in\mathcal{N}}$ denotes the profile of type distributions indexed by participants $\mathcal{N}$. Subsequently, at the second stage, it elicits the realized types $\theta_{\mathcal{N}} := (\theta_i)_{i\in\mathcal{N}}$ where for each $i\in\mathcal{N}$, $\theta_i\sim q_i$, and determines second-stage decisions $y\in\mathcal{Y}_{\mathcal{N}}$ such as reserve activation. Moreover, each participant $i$ receives a payment $t_i^{(1)}$ at the first stage and $t_i^{(2)}$ at the second stage. This leads us to define a two-stage stochastic market, inspired by~\cite{ieong2007stochastic}, as follows. For ease of notation, we mention the subscript only when it is not the full set, and hereafter write $\mathcal{X}:= \mathcal{X}_\mathcal{N}, \mathcal{Y}:= \mathcal{Y}_\mathcal{N}, q:= q_\mathcal{N}, \theta:=\theta_\mathcal{N}$.}

\begin{figure}
    \centering
    \begin{tikzpicture}
        \node[node1, text width=2cm, align=center] (s1) {First stage (time-ahead)};
        \node[node1, text width=2cm, right of = s1, node distance = 4.5cm, align=center] (s2) {Second stage (real-time)};
        \draw[arrow] (s1) -- node[above] (x1) {} (s2);
        \node[node1, below of = x1, node distance=3.35cm, text width=6cm, align=center] (ut) {Participant utility: Payments - Cost\\$u_i = t_i^{(1)} + t_i^{(2)} - c_i(x, y, \theta_i)$ \vspace{4pt} \\ Operator's decision cost: \\ $c^{(1)}(x) + c^{(2)}(x,y)$};
        \draw[arrow] (s1.south) -- node[right, text width=1.5cm, align=center] (t1) {First-stage payment $t_i^{(1)}(q)$} node [left, text width=1.5cm, align=center] {First-stage decision $x^\star(q)$} ([yshift=-1.7cm]s1.south);
        \draw[arrow] (s2.south) -- node[left, text width = 1.9cm, align=center] (t2) {Second-stage decision $y^\star(q, x, \theta)$} node [right, text width=1.9cm, align=center] {Second-stage payment $t_i^{(2)}(q,x,\theta)$} ([yshift=-1.7cm]s2.south);
        \node[above of = s1, text width=2.5cm, align=center, node distance=1.6cm] (q1) {Type distributions \\$(q_i)_{i\in\mathcal{N}}$};
        \draw[arrow] (q1.south) -- ([yshift=-0.4cm]q1.south);
        \node[above of = s2, text width=2cm, align=center, node distance=1.6cm] (theta1) {Realized types \\$(\theta_i)_{i\in\mathcal{N}}$};
        \draw[arrow] (theta1.south) -- ([yshift=-0.4cm]theta1.south);
        \draw[arrow] (q1.east) -- node[above, text width=1.5cm, align=center] {type realization} node[below] {$\theta_i\sim q_i$} (theta1.west);
    \end{tikzpicture}
    \caption{The two-stage stochastic market process.}\label{fig:twostagestochmarket}
\end{figure}
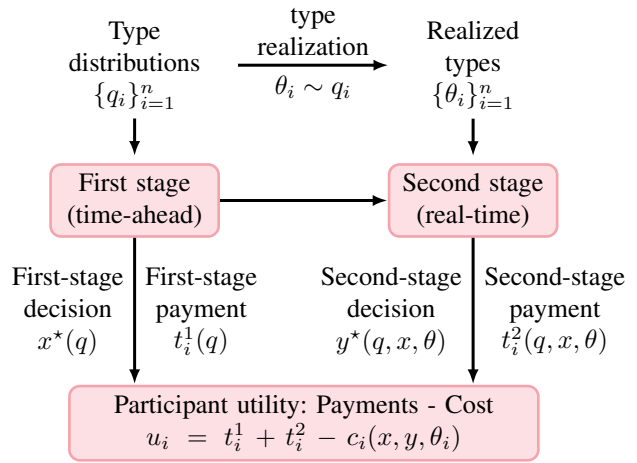

\begin{definition}\label{def:stochmar}
    A \emph{two-stage stochastic market} is defined by a pair of mechanisms, where at each stage, the market operator determines a stage decision and participant payments that depend on the information revealed by the participants up to that stage and the previous stage decisions, as shown in Fig.~\ref{fig:twostagestochmarket}. Specifically,
    \begin{enumerate}
        \item the first stage accepts type distributions $q$ and determines a first-stage decision $x^\star(q)$, with payments $\left(t_i^{(1)}(q)\right)_{i\in\mathcal{N}}$ for all participants, where ${x^\star: \mathbb{Q}^n \rightarrow \mathcal{X}}$ and ${t_i^{(1)}: \mathbb{Q}^n\rightarrow\realN{}}$ for each $i$, depend on the reported type distributions; and
        \item the second stage accepts the realized types $\theta$ and determines the second-stage decision $y^\star(q, x, \theta)$, with payments $\left(t_i^{(2)}(q,x,\theta)\right)_{i\in\mathcal{N}}$ for all participants, where ${y^\star:\mathbb{Q}^n \times \mathcal{X} \times \Theta^n\rightarrow \mathcal{Y}}$ and ${t_i^{(2)}: \mathbb{Q}^n\times\mathcal{X}\times\Theta^n\rightarrow\realN{}}$ for each $i$, depend on the reported type distributions, the first-stage decision, and the reported types.
    \end{enumerate}
\end{definition}

\section{Two-stage stochastic dispatch problem}\label{sec:stochasticdispatch}
In this section, we define the stage decision rules $x^\star$ and $y^\star$ by describing a general two-stage stochastic dispatch problem arising in power system applications, an instance of the stochastic programming frameworks in~\cite{shapiro2014stochastic,bertsekas1976dynamic}.

\subsection{General description}
The goal of the market operator is to determine first- and second-stage decisions $x\in\mathcal{X}$ and $y\in\mathcal{Y}$, to minimize the system cost which is given by
\begin{equation}
    l(x,y,\theta) := c^{(1)}(x) + c^{(2)}(x,y) + \sum_{i\in\mathcal{N}} c_i(x, y, \theta_i),
\end{equation}
where $c^{(1)}:\mathcal{X}\rightarrow\realN{}$ and $c^{(2)}:\mathcal{X}\times\mathcal{Y}\rightarrow\realN{}$ are the costs incurred by the market operator for the first- and second-stage decisions. $c_i:\mathcal{X}\times\mathcal{Y}\times\Theta\rightarrow\realN{}$ denotes the cost incurred by each participant $i$ which depends on both the stage decisions as well as its realized type. \added{Throughout, the index set of a profile is part of the profile itself and is not passed as a separate argument, thus a summation in these definitions ranges over the index set of the profile supplied as its argument, which is $\mathcal{N}$ in the above equation.} 

Due to uncertainty about participant types, the market operator minimizes expected system cost at the first stage, defining the first-stage decision as
\begin{multline}
    x^\star(q)\in \arg\min_{x\in\mathcal{X}}\ l^{(1)}(x, q) := c^{(1)}(x) +  \\
    \underset{\theta\sim q}{\mathbb{E}}\bigg[c^{(2)}(x, y^\star(q, x, \theta)) + \sum_{i\in\mathcal{N}} c_i(x, y^\star(q, x, \theta), \theta_i) \bigg] \label{eq:firstoutcome}
\end{multline}
where $y^\star$ denotes the second-stage decision rule defined below\footnote{assuming unique solutions due to ties broken by a fixed rule.}. $\theta\sim q$ denotes that for each $i\in\mathcal{N}$, $\theta_i \sim q_i$. At the second stage, the market operator further minimizes the system cost based on the first-stage decision and realized participant types. For a given first-stage decision $x$, the second-stage decision problem reads as
\begin{align}
    y^\star(q,x,\theta)\in\arg\min_{y\in\mathcal{Y}(x)}\  & l^{(2)}(x, y, \theta):= \nonumber                                                                                     \\
                                                         & \hspace{-2em} c^{(2)}(x, y) + \sum_{i\in\mathcal{N}} c_i(x, y, \theta_i).\label{eq:secondoutcome}
\end{align}
Note that the feasible space of the second-stage decision may be further constrained by the first-stage decision, ${\mathcal{Y}(x)\subseteq\mathcal{Y}}$.


\subsection{A two-stage stochastic electricity dispatch}\label{sec:elecmarketapp}

Let us consider the market operator's problem in the context of a two-stage electricity market, where the market operator desires to procure a fixed demand $D$ of energy from several participating producers. At the first stage, the market operator elicits type distributions $q$, and optimizes the system cost~\eqref{eq:firstoutcome} which determines $x = \left((\bar{d}_i)_{i\in\mathcal{N}}, \bar{r}, g\right)$ consisting of three types of decisions: 1) unit dispatch $(\bar{d}_i)_{i\in\mathcal{N}}$, where $\bar{d}_i \in \{0,1\}$ is a binary commitment variable, 2) reserve energy capacity $\bar{r}\in\realN{}_+$ which is the amount of reserve energy available to be activated in the second stage, and 3) dispatchable power $g\in\realN{}_+$ which is delivered without uncertainty. Let $\alpha_{\bar{r}}, \alpha_g$ be the marginal costs of reserve capacity and dispatchable power, respectively. Then, the first-stage decision cost is given by
\begin{equation}
    c^{(1)}(x):=\alpha_{\bar{r}} \bar{r} + \alpha_{g} g.
\end{equation}

In the second stage, the producers realize and report their types, and the market operator determines ${y = \left( (d_i)_{i\in\mathcal{N}}, r, s \right)}$ consisting of the volume dispatch $(d_i)_{i\in\mathcal{N}}$, where $d_i \in \realN{}$ denotes the volume to be delivered by participant $i$, reserve activation $r\in\realN{}$, and load shedding $s\in\realN{}_+$ is the amount of unfulfilled load. Let $\alpha_{r}$ and $\alpha_{s}$ be the marginal costs for reserve activation and load shedding, respectively. Then, the second-stage cost is given by
\begin{align}
    c^{(2)}(x, y):= \alpha_{r} \lvert r \rvert + \alpha_{s} s.
\end{align}

The second stage optimizes the system cost~\eqref{eq:secondoutcome}, subject to system constraints, defining the feasible set for the second-stage decision as
\begin{subequations}
    \begin{align}
        \mathcal{Y}(x):=\Big\{  y \ \Big| \ & d_i \in \realN{}, \forall i\in\mathcal{N},\quad r\in\realN{},\quad s\in\realN{}_+                                                  \\
                               & \hspace{-8pt} D - \sum_{i\in\mathcal{N}} \bar{d}_i d_i - g - s = r,\  r \in [-\bar{r}, \bar{r}] \Big\} \label{eq:powerbalance}
    \end{align}
\end{subequations}
where~\eqref{eq:powerbalance} determines the reserve activation required to meet residual demand after considering supply from dispatched producers, dispatchable power, and load shedding, with activation limited by the reserves procured in the first stage.

We consider producers with a three-dimensional type space ${\Theta \subset \realN{3}}$ (i.e., $p=3$), capturing characteristics like production capacity and flexibility costs. The corresponding cost function reads as,
\begin{equation}
    c_i(x, y, \theta_i):= \begin{cases}
        \theta_{i,1} (\theta_{i,2} - d_i), & d_i \leq \theta_{i,2} \\
        \theta_{i,3} (d_i - \theta_{i,2}), & d_i > \theta_{i,2},
    \end{cases}
\end{equation}
which models a non-dispatchable power producer with an uncertain production capacity, and ability to adjust their production at a cost. Specifically, $\theta_{i,2}$ represents their baseline production, and $\theta_{i,1},\theta_{i,3}$ represent the marginal costs of down and upregulation, respectively. For a non-dispatchable producer without any storage asset, upregulation is not feasible and thus $\theta_{i,3}\rightarrow \infty$, however, for hybrid assets like wind power with attached storage, upregulation is feasible, albeit at a cost. For the set of type distributions $\mathbb{Q}$, we consider a family of Gaussian distributions, defined as
\begin{equation}
    \mathbb{Q}:= \left\{\mathrm{N}(\mu, \Sigma)\ |\ \mu \in\realN{3}, \Sigma\in \mathbb{S}^3_+ \right\},
\end{equation}
where $\mathbb{S}^3_+$ denotes the set of all $3\times 3$ symmetric positive semidefinite matrices. Thus, each $q_i\in\mathbb{Q}$ corresponds to a Gaussian distribution with mean $\mu_i$ and covariance $\Sigma_i$.

\section{Payment rule and market properties}\label{sec:mechd}


In competitive markets, each participant aims to maximize its utility $u_i = t_i^{(1)} + t_i^{(2)} - c_i(x, y, \theta_i)$, defined as the difference between the received payments and the incurred cost. As discussed in the previous section, the stage decisions and payments are determined based on the participants' reported type information. Consequently, participants may misreport type information strategically to influence stage decisions and payments to maximize their utility. While such strategic behavior may maximize the participant utility, it increases total system cost due to misreporting of type information. To this end, a main challenge is to ensure that participants' reported information is truthful, thereby ensuring cost efficiency of the stage decisions.

Depending on the payment functions, the utility maximizing strategy of a participant may or may not align with truthful reporting. The property of a mechanism to incentivize truthful reporting of private information is referred to as \emph{incentive compatibility}. For single-stage mechanisms, there are several notions of incentive compatibility in the literature, which differ in the extent of knowledge a participant needs about the others' types for truthful reporting to be optimal~\cite{nisan2007algorithmic}. For instance, the strongest notion is \ac{DSIC}, where truth telling is optimal irrespective of others' strategy, while in an ex-post incentive-compatible mechanism, truth telling is optimal when others also report truth.

For the two-stage market setting, we need a sequential generalization of these classical solution concepts, since the payment structure needs to align incentive at each stage. Consider \emph{sequential ex-post incentive compatibility}~\cite{ieong2007stochastic}\footnote{\ac{DSIC} is not achievable in the two-stage stochastic market setting.}, where a participant
\begin{enumerate}
    \item cannot improve its expected utility by misreporting type distribution in the first stage, given that other participants report truthfully; and
    \item cannot improve its utility by misreporting its realized type in the second stage, given that other participants report truthfully.
\end{enumerate}

Incentive compatibility renders truth telling an optimal strategy, however, the utility received by a participant on truthful reporting may still be negative, which discourages market participation altogether. \emph{Individual rationality} refers to the market property where each participant is guaranteed a non-negative utility, meaning that a participant is always better off participating than not. In the two-stage setting, we consider a mechanism to be ex-ante individually rational, if the expected utility at first stage is non-negative. This is a weaker notion in contrast to ex-post individual rationality, where utility is guaranteed to be non-negative for each type realization.
\ac{VCG} payments are \ac{DSIC} and individually rational in the single-stage setting, motivating their extension to the two-stage setting.

Recall the \ac{VCG} payments for a single-stage mechanism~\cite{nisan2007algorithmic}. Consider a single-stage system cost minimization problem for a set of participants $\mathcal{N}$ with types $\theta:=(\theta_i)_{i\in\mathcal{N}}$,
\begin{equation*}
    f(\theta) \in \arg\min_{y\in\mathcal{Y}}\ l(y, \theta) := c(y) + \sum_{i\in\mathcal{N}} c_i(y, \theta_i),
\end{equation*}
where $c$ and $c_i$ are reused for their single-stage counterparts, denoting the market decision and participant $i$ cost. Then, for each participant $i$, the \ac{VCG} payment is
\begin{equation*}
    c_i(y, \theta_i) - l(f(\theta), \theta) + l\left(f(\theta_{-i}), \theta_{-i}\right),
\end{equation*}
where $\theta_{-i} := \theta_{\mathcal{N}\setminus\{i\}}$ denotes the type profile excluding participant $i$, and $f\left(\theta_{-i}\right)$ denotes the market decision without participant $i$. Intuitively, the payment is a sum of the participant's cost (first-term) and its marginal contribution to the system cost (last two terms). This aligns the individual utility maximization problem with system cost minimization, incentivizing truthful reporting. 

We extend the \ac{VCG} payments to the two-stage setting by defining stage payments as follows. Consider the first-stage payment function:
\begin{equation}
    t_i^{(1)}(q) = -c^{(1)}(x)+ c^{(1)}(x_{-i}), \label{eq:firstpayments}
\end{equation}
where $x = x^\star(q)$, $x_{-i} = x^\star(q_{-i})$ are the optimal first-stage decisions with and without participant $i$, where ${q_{-i} = q_{\mathcal{N}\setminus\{i\}}}$ denotes the family of other participants' reported type distributions. The first-stage payment is equal to the decrease in the first-stage decision cost due to participant $i$. Further, consider the second-stage payment function
\begin{multline}
    t_i^{(2)}(q, x, \theta) = c_i(x, y, \theta_i) - l^{(2)}(x, y, \theta) + \\l^{(2)}(x_{-i}, y_{-i}, \theta_{-i}), \label{eq:secondpayments}
\end{multline}
where $y = y^\star(q, x,\theta)$ denotes the optimal second-stage decision, given the first-stage decision and the reported types, and $y_{-i} = y^\star(q_{-i}, x_{-i}, \theta_{-i})$ denotes the optimal second-stage decision without participant $i$. The payment is a sum of the participant cost (first-term), and its marginal contribution to second-stage system cost (last two terms).

\begin{theorem}\label{thm:1}
    Consider the two-stage stochastic market mechanism in Definition~\ref{def:stochmar}, with stage decisions given by Eqs.~\eqref{eq:firstoutcome},~\eqref{eq:secondoutcome} and payments given by Eqs.~\eqref{eq:firstpayments},~\eqref{eq:secondpayments}. Then, the mechanism is
    \begin{enumerate}
        \renewcommand{\labelenumi}{\Alph{enumi})}
        \item sequentially ex-post incentive-compatible; and
        \item ex-ante individually rational.
    \end{enumerate}
\end{theorem}
Moreover, incentive compatibility of reporting true type information ensures the cost efficiency of the implemented stage decisions, and hence the mechanism is efficient. The proof of the above theorem is included in the Appendix.


\section{Numerical experiment}\label{sec:numexp}
In this section, we numerically evaluate the proposed mechanism in the context of an electricity market defined in Section~\ref{sec:elecmarketapp}. The first-stage decision problem~\eqref{eq:firstoutcome} is formulated as a standard two-stage stochastic program with recourse decisions (second-stage decisions). We use a finite sample approximation for the expected second-stage cost with a sample size of 1000, and formulate the problem as a mixed-integer linear program. Both stage problems are solved using Gurobi~\cite{gurobi}. The computational time for solving the first-stage decision is approximately one second, while the second-stage problem requires negligible time. All the experiments are run on an Apple M3 8-core CPU with 16GB of RAM, and the corresponding code is available online\footnote{Code available at: https://github.com/EMA-DTU/STEM}.

We consider $n=5$ non-dispatchable power producers, each characterized by an uncertain production with an expected volume of $\mu_{i,2} = 20$ MWh. Each producer has a distinct production variance, and down and upregulation costs. We take the up- and downregulation costs to be deterministic, so that the only non-zero entry of participant $i$'s covariance matrix $\Sigma_i$ is $\Sigma_{i,2,2}$, the variance of the forecast production $\theta_{i,2}$. The market operator's demand is $100$ MWh, and it incurs $\alpha_{\bar{r}}=10$ \texteuro{}/MWh for reserve procurement, $\alpha_{g}=6$ \texteuro{}/MWh for dispatchable power, $\alpha_{r}=8$ \texteuro{}/MWh for reserve activation, and $\alpha_{s}=200$ \texteuro{}/MWh for load shedding.

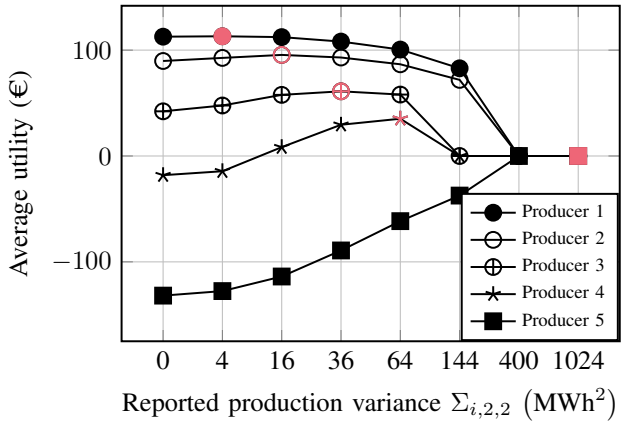
\begin{figure}[t]
    \centering
    \begin{tikzpicture}
        \begin{axis}[
                grid,
                xticklabel style={rotate=0},
                ymin = -175,
                height = 0.68\linewidth,
                width = 0.95\linewidth,
                mark size = 3,
                table/col sep=comma,
                xtick = {1,2,3,4,5,6,7,8},
                xticklabels = {0, 4, 16, 36, 64, 144, 400, 1024},
                ylabel = {Average utility (\texteuro{})},
                xlabel = {Reported production variance $\Sigma_{i,2,2}$ $\left(\text{MWh}^2\right)$},
                line width=0.75,
                legend style = {font=\scriptsize, legend cell align=left, name=a, xshift=0.12cm, yshift=-2.22cm},
            ]
            \addplot[name path = up, mark = *, opacity=1] table[x index=0, y=a] {ic.csv};
            \addlegendentry{Producer 1};

            \addplot[name path = up, mark = o, opacity=1] table[x index=0, y=b] {ic.csv};
            \addlegendentry{Producer 2};

            \addplot[name path = up, mark = oplus, opacity=1] table[x index=0, y=c] {ic.csv};
            \addlegendentry{Producer 3};

            \addplot[name path = up, mark = star, opacity=1] table[x index=0, y=d] {ic.csv};
            \addlegendentry{Producer 4};

            \addplot[name path = up, mark = square*, opacity=1] table[x index=0, y=e] {ic.csv};
            \addlegendentry{Producer 5};
            \addplot[mark=*, red] coordinates {(2, 113.07)};
            \addplot[mark=o, red] coordinates {(3, 95.47)};
            \addplot[mark=oplus, red] coordinates {(4, 61.09)};
            \addplot[mark=star, red] coordinates {(5, 35.19)};
            \addplot[mark=square*, red] coordinates {(8, 0)};
        \end{axis}
    \end{tikzpicture}
    \caption{Producer average utility as a function of the reported production variance; red markers indicate true variance.}\label{fig:IC}
\end{figure}

Fig.~\ref{fig:IC} shows each producer's average utility against its reported production variance. Since every producer maximizes its utility by reporting true variance, it validates the incentive compatibility of the mechanism.

\begin{figure}
    \centering
    \begin{tikzpicture}
        \begin{axis}[
                width = 0.9\linewidth,
                height = 0.4\linewidth,
                xtick = {1,2,3,4,5},
                xticklabels = {4, 16, 36, 64, 1024},
                ybar stacked,
                bar width=10pt,
                ylabel={Payment (\texteuro{})},
                xlabel = Production variance $\Sigma_{i,2,2}$ $\left(\text{MWh}^2\right)$,
                xtick=data,
                table/col sep=comma,
                legend columns = -1,
                legend style = {font=\footnotesize, draw=none, legend cell align=left, fill=none, name=a, xshift=1cm, yshift=0.9cm},
                ymin=-50,
                ymax=150,
                name=top,
            ]
            \draw [dotted] (0,0) -- (6,0);
            \addplot[
                draw=red,
                fill=red!40,
            ] table[
                    x index=0,
                    y = secondpayments,
                ] {payments_unc.csv};
            \addlegendentry{Second-stage payments}

            \addplot[fill=cyan!40, draw=cyan] table[x index=0, y=firstpayments] {payments_unc.csv};
            \addlegendentry{First-stage payments}

            \addlegendimage{legend image code/.code={
                        \node[black, fill=black, circle, inner sep=1.5pt] at (0.0cm,0cm) {};
                    }}
            \addlegendentry{Total payments}
        \end{axis}
        \begin{axis}[
                width = 0.9\linewidth,
                height = 0.4\linewidth,
                axis x line=none,
                axis y line=none,
                ymin=-50,
                ymax=150,
                xtick=\empty,
                ytick=\empty,
                table/col sep=comma,
            ]

            \addplot[black,
                mark=*,
                error bars/.cd,
                y dir=both,
                y explicit] table[x index=0, y expr=\thisrow{firstpayments} + \thisrow{secondpayments}, y error expr=\thisrow{secondpaymentsstd}] {payments_unc.csv};

            \node[anchor = west] at (0.7, -30) {\footnotesize Increasing uncertainty $\rightarrow$};
        \end{axis}
        \begin{axis}[
                anchor = north,
                yshift = -1.2cm,
                xtick = {1,2,3,4,5},
                xticklabels = {2, 5, 10, 15, 20},
                at = {(top.south)},
                width = 0.9\linewidth,
                height = 0.4\linewidth,
                ybar stacked,
                bar width=10pt,
                ylabel={Payment (\texteuro{})},
                xlabel={Downregulation cost $\theta_{i,1}$ (\texteuro{}/MWh)},
                xtick=data,
                table/col sep=comma,
                ymin=-80,
                ymax=250,
            ]
            \draw [dotted] (0,0) -- (6,0);
            \addplot[
                draw=red,
                fill=red!40,
            ] table[
                    x index=0,
                    y = secondpayments,
                ] {payments_flex.csv};

            \addplot[fill=cyan!40, draw=cyan] table[x index=0, y=firstpayments] {payments_flex.csv};
        \end{axis}
        \begin{axis}[
                at = {(top.south)},
                yshift = -1.2cm,
                anchor = north,
                width = 0.9\linewidth,
                height = 0.4\linewidth,
                axis x line=none,
                axis y line=none,
                ymin=-80,
                ymax=250,
                legend style = {font=\footnotesize, at = (a.south west), yshift=0.15cm, xshift=-0.42cm, anchor=north west, draw=none, fill=none},
                xtick=\empty,
                ytick=\empty,
                table/col sep=comma,
            ]

            \addplot[black,
                mark=*,
                error bars/.cd,
                y dir=both,
                y explicit] table[x index=0, y expr=\thisrow{firstpayments} + \thisrow{secondpayments}, y error expr=\thisrow{secondpaymentsstd}] {payments_flex.csv};

            \node[anchor = west, yshift=0.15cm] at (0.7, -60) {\footnotesize Increasing flexibility cost $\rightarrow$};
        \end{axis}
    \end{tikzpicture}
    \caption{Average payments to each producer at both the stages, where the error bars represent standard deviation due to uncertainty of realized production in the second stage.}\label{fig:payments}
\end{figure}
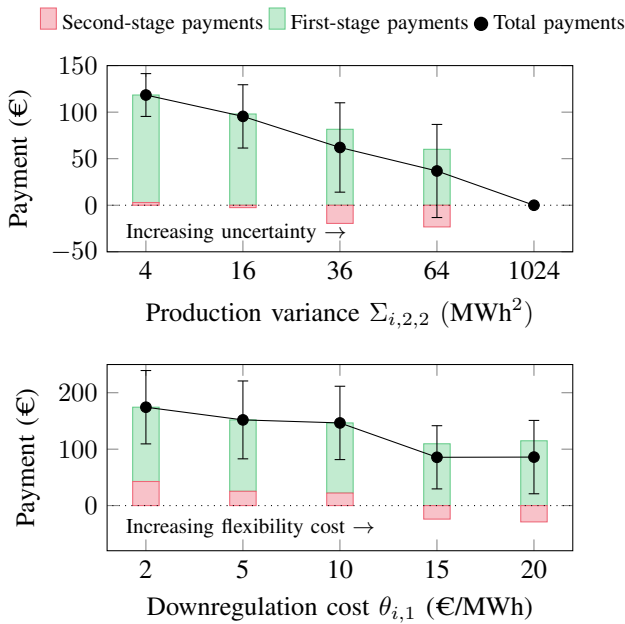

Fig.~\ref{fig:payments} shows the impact of a producer's production uncertainty (top) and flexibility costs (bottom) on the total payment. In each case, other parameters are kept identical. The payments reduce with an increase in production uncertainty due to a higher contribution to the system cost. Further, higher flexibility costs reduce the utilization of the producer's flexibility relative to other producers, lowering its contribution to system cost reduction and, consequently, its payments. These results illustrate that the pricing accounts not just for volume, but also for flexibility and uncertainty. In doing so, it generates an explicit price signal for flexible assets, generating financial incentives and facilitating their efficient integration into system operation and dispatch.

\begin{table}[]
    \centering
    \caption{Comparison of the proposed mechanism with a deterministic mechanism emulating current market practice.}\label{tab:marketoutcome}
    \begin{tabular}{c|c|c|c}
                            & Stochastic               & Deterministic                       & Deterministic                     \\
                            & (proposed)               & $\hat{\Sigma}_{i,2,2} = 100$ & $\hat{\Sigma}_{i,2,2} = 4$ \\
        \hline
        Dispatchable ($g$)        & 19 MWh                   & 0 MWh                               & 0 MWh                             \\
        Reserve capacity ($\bar{r}$)   & 20 MWh                   & 35 MWh                              & 8 MWh                             \\
        Average system cost & \textbf{416} \texteuro{} & 1162 \texteuro{}                    & 471 \texteuro{}
    \end{tabular}
\end{table}

Table~\ref{tab:marketoutcome} compares the proposed market mechanism with a deterministic market that emulates the current market process, where producers report only the expected production and the market operator estimates production variance. We consider two cases, where the market operator overestimates ${(\hat{\Sigma}_{i,2,2}=100)}$ and underestimates ${(\hat{\Sigma}_{i,2,2}=4)}$ production variance for each producer. The producers in this study correspond to those in Fig.~\ref{fig:payments} (top). The proposed mechanism does not dispatch the fifth producer due to a high production uncertainty (${\Sigma_{5,2,2}=1024}$), and dispatchable power of $19$ MWh is procured instead. This highlights the mechanism's effectiveness to consider the cost-uncertainty tradeoff while dispatching energy assets. In contrast, the deterministic market unaware of producer-specific uncertainty, dispatches all the producers, resulting in suboptimal first-stage decisions, increasing system cost.


\section{Conclusion}\label{sec:conc}
We considered a two-stage stochastic market mechanism, which elicits forecasted asset characteristics including production distribution at the first stage, and realized production at the second stage, and pays participants based on their impact on the system cost due to their flexibility and uncertainty. The mechanism is sequentially ex-post incentive-compatible, and ex-ante individually rational. As a result, the market operator receives truthful information regarding each producer's uncertainty and flexibility characteristics, enabling efficient dispatch and reducing total system cost. Finally, we validated the proposed theoretical claims using an electricity market case study, showing the effectiveness of the proposed mechanism to minimize system cost and generate incentives for participants to report their best forecast.

Two main limitations of the proposed mechanism merit further investigation: 1) the reliance on a relatively weak notion of ex-post incentive compatibility at the first-stage, leaving it susceptible to collusion; and 2) computational complexity of the pricing rule as the two-stage stochastic dispatch is solved $n$ times to evaluate individual contributions for payment calculation.

\section*{Acknowledgment}
The authors would like to thank Spyros Chatzivasileiadis (DTU) for valuable comments and suggestions that significantly improved the quality of this work.

\bibliographystyle{ieeetr}
\bibliography{bibliography}

\appendix

\section*{Proof of Theorem~\ref{thm:1}}\label{app:proof}

\subsection{Sequential ex-post incentive compatibility}
For sequential ex-post incentive compatibility, we need to show incentive compatibility at each stage. We first show that the second-stage payments induce a \ac{DSIC} mechanism. Given second-stage payments~\eqref{eq:secondpayments}, the utility maximization problem for participant $i$ at the second stage reads as
\begin{subequations}
    \begin{align}
        \max_{\hat{\theta}_i}\quad & - c_i(x, y, \theta_i) + t_i^{(2)}(q, x, \hat{\theta}) + t_i^{(1)}(q) \\ 
        \text{s.t.}\quad           & y = y^\star(q, x, \hat{\theta}),
    \end{align}
\end{subequations}
where $\hat{\theta}:=(\hat{\theta}_i)_{i\in\mathcal{N}}$ represents type reports from all the participants which might be different than the true types $\theta$. The cost of the participant depends on its realized type $\theta_i$, while the payment and the second-stage decision depend on the reported type. Expanding, and eliminating terms independent of the second-stage report $\hat{\theta}_{i}$, we get
\begin{subequations}
    \begin{align}
        \max_{\hat{\theta}_i}\ & - c_i(x, y, \theta_i) - c^{(2)}(x, y) - \hspace{-6pt} \sum_{j\in\mathcal{N}\setminus\{i\}} c_j(x, y, \hat{\theta}_j) \label{eq:utilsecondstage} \\
        \text{s.t.}\           & y = y^\star(q, x, \hat{\theta})
    \end{align}
\end{subequations}

Note that~\eqref{eq:utilsecondstage} is equal to $-l^{(2)}\left(x, y, (\theta_i, \hat{\theta}_{-i})\right)$, \added{where $(\theta_i, \hat{\theta}_{-i})$ represents the type profile obtained from $\hat{\theta}$ by replacing the $i$-th element by $\theta_i$.} Since $y^\star\left(q, x, (\theta_i, \hat{\theta}_{-i})\right)$ is a minimizer of $l^{(2)}\left(x, \cdot, (\theta_i, \hat{\theta}_{-i})\right)$, and $y^\star(q, x, \hat{\theta}) \in \mathcal{Y}(x)$ (i.e., a feasible solution of~\eqref{eq:secondoutcome}), $\hat{\theta}_i=\theta_i$ is an optimal report. Note that, truthful reporting remains optimal irrespective of the other participants' reports $\hat{\theta}_{-i}$, achieving a stronger notion of \ac{DSIC} in second stage.

Now, given the first-stage payments~\eqref{eq:firstpayments}, the expected utility for participant $i$ at first stage is given by
\begin{subequations}
    \begin{align}
        \max_{\hat{q}_i}\  & t^{(1)}_i(\tilde{q}) + \mathbb{E}_{\theta\sim q} \big[ t_i^{(2)}(\tilde{q}, x, \theta) - c_i(x, y_\theta, \theta_i)\big] \\
        \text{s.t.}\       & x = x^\star(\tilde{q})                                                                                           \\
                           & y_\theta = y^\star(q, x, \theta),\ \forall \theta \in \Theta^n,
    \end{align}
\end{subequations}
where $\tilde{q}:=(\hat{q}_i, q_{-i})$ represents a false report from participant $i$ while all other participants report truthfully. The payments and the first-stage decision depend on the reported type distribution, while the types in second stage are realized according to the true type distributions. Due to the incentive-compatibility of the second stage, the second-stage payment depends on the true realized type. Expanding the payments and eliminating terms that are independent of $\hat{q}_i$, we get
\begin{subequations}
    \begin{align}
        \max_{\hat{q}_i}\  & - c^{(1)}(x) - \mathbb{E}_{\theta\sim q} \left[ l^{(2)}(x, y_\theta, \theta) \right] \label{eq:utilfirststage} \\
        \text{s.t.}\       & x = x^\star(\tilde{q})                                                                      \\
                           & y_\theta = y^\star(q, x, \theta),\ \forall \theta\in\Theta^n
    \end{align}
\end{subequations}

Note that~\eqref{eq:utilfirststage} is equal to $-l^{(1)}(x, q)$. Since $x^\star(q)$ is a minimizer of $l^{(1)}(\cdot,q)$, and $x^\star(\tilde{q}) \in \mathcal{X}$, $\hat{q}_i = q_i$ is an optimal report. Thus, truth telling when other participants report truth is an equilibrium strategy, and the proposed mechanism is sequentially ex-post incentive-compatible.

\subsection{Individual rationality}
Now, we show that each participant's expected utility at the first stage is non-negative. For the given payments, the expected utility at the first stage for participant $i$ is given by:
\begin{equation}
    \mathbb{E}_{\theta\sim q}\left[t_i^{(1)}(q) + t_i^{(2)}(q, x, \theta) - c_i(x, y, \theta_i)\right]
\end{equation}
Expanding the terms, we get
\begin{multline}
    \mathbb{E}_{\theta\sim q} \Big[c^{(1)}(x_{-i}) -c^{(1)}(x) + l^{(2)}(x_{-i}, y_{-i}, \theta_{-i}) \\
        - l^{(2)}(x, y, \theta) + c_i(x, y, \theta_i) - c_i(x, y, \theta_i)\Big],
\end{multline}
where $x_{-i} = x^\star(q_{-i})$, $y_{-i} = y^\star(q_{-i}, x_{-i}, \theta_{-i})$. Simplifying further, we get
\begin{equation}
     l^{(1)}(x_{-i}, q_{-i}) - l^{(1)}(x, q).
\end{equation}
Since $x_{-i}$ with $\bar{d}_i = 0$ is a feasible solution, and $x^\star(q)$ is a minimizer of $l^{(1)}(\cdot, q)$, ${l^{(1)}(x, q) \leq l^{(1)}(x_{-i}, q_{-i})}$, and thus, the expected utility is non-negative.

\end{document}

%% file: acronyms.tex
\newcommand{\realN}[1]{
    \mathbb{R}^{#1}
}

\DeclareAcronym{VCG}{
    short = VCG,
    long = Vickrey-Clarke-Groves,
    tag = abbrev
}

\DeclareAcronym{WPP}{
    short = WPP,
    long = wind power producer,
    tag = abbrev
}

\DeclareAcronym{DSIC}{
    short = DSIC,
    long = dominant strategy incentive-compatible,
    tag = abbrev
}

%% file: main_latest.bbl
\begin{thebibliography}{10}

\bibitem{bjorndal2018challenge}
E.~Bjørndal, M.~Bjørndal, K.~Midthun, and A.~Tomasgard, ``Stochastic electricity dispatch: A challenge for market design,'' {\em Energy}, vol.~150, pp.~992--1005, May 2018.

\bibitem{morales2012pools}
J.~M. Morales, A.~J. Conejo, K.~Liu, and J.~Zhong, ``Pricing electricity in pools with wind producers,'' {\em IEEE Transactions on Power Systems}, vol.~27, pp.~1366--1376, Aug. 2012.

\bibitem{pritchard2010single}
G.~Pritchard, G.~Zakeri, and A.~Philpott, ``A single-settlement, energy-only electric power market for unpredictable and intermittent participants.,'' {\em Operations Research}, vol.~58, pp.~1210--1219, July 2010.

\bibitem{zavala2017consistent}
V.~M. Zavala, K.~Kibaek, M.~Anitescu, and J.~Birge, ``A stochastic electricity market clearing formulation with consistent pricing properties.,'' {\em Operations Research}, vol.~65, pp.~557--576, May 2017.

\bibitem{kazempour2018scenario}
J.~Kazempour, P.~Pinson, and B.~F. Hobbs, ``A stochastic market design with revenue adequacy and cost recovery by scenario: Benefits and costs,'' {\em IEEE Transactions on Power Systems}, vol.~33, pp.~3531--3545, July 2018.

\bibitem{mays2021quasi}
J.~Mays, ``Quasi-stochastic electricity markets.,'' {\em INFORMS Journal on Optimization}, vol.~3, pp.~350--372, Oct. 2021.

\bibitem{morales2014improved}
J.~M. Morales, M.~Zugno, S.~Pineda, and P.~Pinson, ``Electricity market clearing with improved scheduling of stochastic production,'' {\em European Journal of Operational Research}, vol.~235, pp.~765--774, June 2014.

\bibitem{dvorkin2020chance}
Y.~Dvorkin, ``A chance-constrained stochastic electricity market,'' {\em IEEE Transactions on Power Systems}, vol.~35, pp.~2993--3003, July 2020.

\bibitem{dvorkin2019asymmetry}
V.~Dvorkin, J.~Kazempour, and P.~Pinson, ``Electricity market equilibrium under information asymmetry,'' {\em Operations Research Letters}, vol.~47, pp.~521--526, Nov. 2019.

\bibitem{tang2015random}
W.~Tang and R.~Jain, ``Market mechanisms for buying random wind,'' {\em IEEE Transactions on Sustainable Energy}, vol.~6, pp.~1615--1623, Oct. 2015.

\bibitem{papakonstantinou2016information}
A.~Papakonstantinou and P.~Pinson, ``Information uncertainty in electricity markets: Introducing probabilistic offers,'' {\em IEEE Transactions on Power Systems}, vol.~31, pp.~5202--5203, Nov. 2016.

\bibitem{koumpis2026restoring}
N.~Koumpis, K.~Kar, L.~Tassiulas, and M.~Zampetakis, ``Restoring incentive compatibility in two-stage energy markets with prosumers,'' no.~arXiv:2606.25910, 2026.

\bibitem{hubner2026stochastic}
T.~Hübner, ``On the design of stochastic electricity auctions,'' Apr. 2026.

\bibitem{ieong2007stochastic}
S.~Ieong, A.~Man-Cho~So, and M.~Sundararajan, {\em Stochastic Mechanism Design}, vol.~4858, pp.~269--280.
\newblock Berlin, Heidelberg: Springer Berlin Heidelberg, 2007.

\bibitem{shapiro2014stochastic}
A.~Shapiro, D.~Dentcheva, and A.~Ruszczynski, {\em Lectures on Stochastic Programming: Modeling and Theory, Second Edition}.
\newblock USA: Society for Industrial and Applied Mathematics, 2014.

\bibitem{bertsekas1976dynamic}
D.~P. Bertsekas, ``Dynamic programming and stochastic control,'' {\em Mathematics in science and engineering}, vol.~125, pp.~222--293, 1976.

\bibitem{nisan2007algorithmic}
N.~Nisan, T.~Roughgarden, E.~Tardos, and V.~V. Vazirani, {\em Algorithmic game theory}.
\newblock Cambridge university press, 2007.

\bibitem{gurobi}
{Gurobi Optimization, LLC}, ``{Gurobi Optimizer Reference Manual},'' 2026.

\end{thebibliography}
